\documentclass[11pt]{article}
\usepackage[margin=1in]{geometry}
\usepackage[T1]{fontenc}
\usepackage{lmodern,amsmath,amssymb,graphicx,booktabs,microtype}
\usepackage[colorlinks=true,allcolors=blue]{hyperref}
\usepackage{caption,longtable}
\usepackage{cite}
\title{A Metamaterial Agentic Swarm: Coordinated Search for Discrete Mechanical Design}
\author{Achuth Chandrasekhar\textsuperscript{1} \and Alexander Zhao\textsuperscript{3} \and Amir Barati Farimani\textsuperscript{1,2,3,4,*}}
\date{\small
\textsuperscript{1}Department of Mechanical Engineering\\
\textsuperscript{2}Department of Chemical Engineering\\
\textsuperscript{3}Department of Materials Science and Engineering\\
\textsuperscript{4}Machine Learning Department\\
Carnegie Mellon University, 5000 Forbes Avenue, Pittsburgh, PA 15213, USA\\[4pt]
\textsuperscript{*}Corresponding author: \href{mailto:barati@cmu.edu}{barati@cmu.edu}}

\DeclareRobustCommand{\agenticon}[1]{\raisebox{-0.25em}{\includegraphics[height=1.3em]{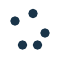}}\hspace{0.45em}}
\begin{document}
\maketitle
\begin{abstract}
Designing discrete metamaterial assemblies is challenging because local unit-cell choices collectively determine the global elastic response. We present Metamaterial Agentic Swarm, an island-based portfolio of forty specialized numerical search agents for unit-cell selection in a pixel-resolved mechanical cloak. Five cooperating islands each contain eight agents responsible for global sampling, spatial proposals, mutation, learning, recombination, diversity, neighborhood exploration, and polishing. Islands maintain separate parents while exchanging evaluated assignments through an elite archive and recombination. The objective is the relative exterior displacement mismatch on a supplied observation mask. Candidate designs are scored by the mismatch in exterior displacement, computed with a symmetry-reduced finite-element model at full pixel resolution. Starting from random assignments within the supplied candidate libraries, the swarm achieves final displacement losses of 0.100590, 0.139274, and 0.135858 at a retrospectively selected reporting endpoint of 961 charged queries per run. Seed~0 achieved the lowest swarm loss, $L=0.100590$, which is 8.72\% below the comparison threshold of 0.1102; the other two runs did not reach this threshold. While they do not establish comparative efficiency or causal benefits from coordination, the results represent a first step towards using agentic swarms for large discrete assembly problems.
\end{abstract}
\noindent\textbf{Keywords:} metamaterial; elastic cloaking; multi-agent search.

\section{Introduction}
Heterogeneous metamaterials realize structural functions by arranging geometrically distinct unit cells. One application is aperiodic mechanical cloaking, in which unit cells are arranged to suppress the displacement disturbance caused by an internal inclusion \cite{wang2022}. The assembly is a global mechanical problem: a locally plausible cell assignment need not preserve the displacement field of the complete structure. Consequently, locally suitable unit cells may produce poor performance when assembled.

A common approach is library-based design, which separates the generation or representation of unit-cell geometries from the selection of their spatial arrangement. Shape descriptors and generative representations have been used to organize mechanical design spaces \cite{shapedna,wang2020}, while diversity-aware dataset selection addresses the coverage of geometry and property spaces \cite{metaset}. Compatible microstructure realization is also central to elastic textures and printable material design \cite{panetta,schumacher}. These contributions motivate the categorical assembly setting; the present work retains the supplied candidate sets rather than generating a new library or reproducing an upstream generative pipeline.

Because evaluating the assembled response is computationally expensive, the search must balance exploration and local refinement. Clockwork combines eight proposal specialties within each of five islands, with separate island parents and a shared archive of evaluated designs. Algorithm portfolios provide a general rationale for combining complementary strategies \cite{xu2010}; they do not guarantee that a larger operator population is more efficient for a particular physical objective. Here, an agent is a stateful numerical proposal operator. No language-model decisions enter the recorded search loop.

\begin{figure}[!ht]
\centering\includegraphics[width=\linewidth]{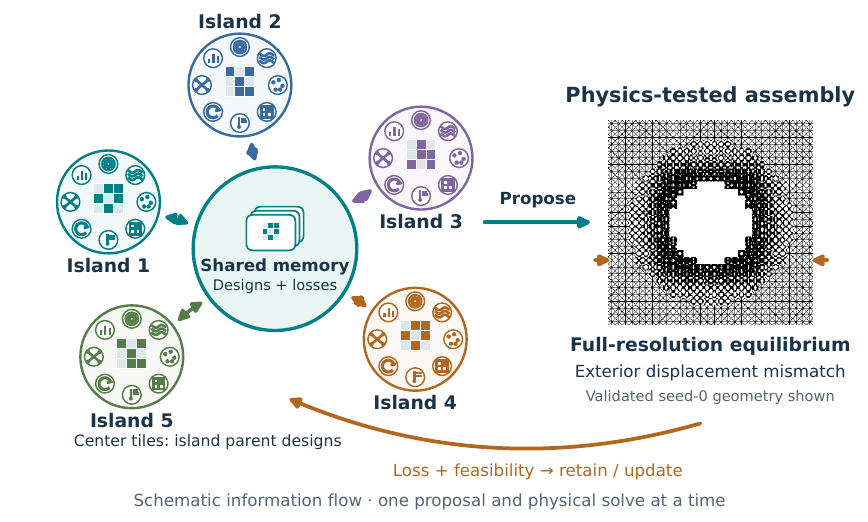}
\caption{Specialization, sharing, and refinement through physical feedback. Forty specialists form five cooperating islands, each with eight icon-coded roles around a parent design (center tile). Shared elite designs and losses support learning, recombination, and incumbent polishing. Physical evaluation returns displacement loss and feasibility feedback for retention and further proposals. Parent tiles and exchange paths are schematic; the assembly at right is the strictly validated seed-0 endpoint. Proposals and physical solves execute serially. Figure~\ref{fig:clockwork} details the implementation.}
\label{fig:concept}
\end{figure}

The framework divides search responsibilities across specialized agents while retaining five distinct parent trajectories for the globally coupled assembly. We evaluate proposed assemblies at full pixel resolution and verify the retained endpoints using fresh equilibrium solves through three random-start runs at an exact reporting budget. Geometry certification, strict endpoint solves, and displacement and stress recovery support the retained designs. Figure~\ref{fig:clockwork} details the information paths. The evidence establishes feasibility and numerical consistency on the cloaking case; it does not establish comparative efficiency or isolate the effects of communication.

\clearpage
\section{Related work and design foundations}
\subsection{Mechanical architecture and inverse design}
Mechanical metamaterials use architecture to control effective properties and structural response \cite{bertoldi2017}. Homogenization-based topology optimization \cite{bendsoe1988} and inverse homogenization \cite{sigmund1994} established systematic routes from target behavior to material layout or unit-cell topology. Combinatorial assembly adds a different design question: how should discrete building blocks be arranged to realize a compatible collective response? Textured mechanical metamaterials demonstrate the importance of this assembly logic \cite{coulais2016}. Clockwork addresses a restricted version of that question, selecting among supplied cell candidates while resolving their assembled pixel geometry.

Mechanical cloaking has been demonstrated through pentamode architectures \cite{buckmann2014} and direct lattice transformations \cite{buckmann2015}. The data-driven aperiodic cloak of Wang et al.\ \cite{wang2022} provides the immediate mechanical setting for the present supplied-library study. Bayesian learning also offers a route to efficient metamaterial exploration \cite{bessa2019}. These precedents motivate the application but do not provide performance baselines here: their design spaces, constitutive assumptions, and evaluation protocols differ. The present evidence concerns three searches in one fixed categorical assembly problem.

\subsection{Roles, shared knowledge, and numerical agency}
Classical agent research distinguishes the capabilities of an autonomous component from the organization of cooperating components \cite{wooldridge1995}. Role-oriented methodologies such as Gaia make responsibilities and interactions explicit \cite{gaia2000}; blackboard systems coordinate specialized knowledge sources through shared problem-solving state \cite{corkill1991}. These are useful architectural precedents for the role map and elite archive in Clockwork. Their connection is at the level of modular responsibilities and shared evidence, rather than an implementation of Gaia's formal models or a general-purpose blackboard reasoning system.

The eight roles are \emph{knowledge-informed}: they encode designer-chosen priors about exploration, spatial correlation, transfer, and local refinement. They do not consult a symbolic knowledge base, infer constitutive laws, or reason in natural language. Each candidate is evaluated by solving the finite-element equilibrium equations, after which the optimizer applies its acceptance rule to the resulting loss. Agent identity specifies persistent state and a proposal responsibility, while the common scheduler determines when that policy acts. Thus the architecture makes its design assumptions inspectable without attributing unimplemented cognitive capabilities to the agents.

\subsection{Cooperative optimization and operator portfolios}
Island-model evolutionary search motivates maintaining distinct trajectories that can exchange information \cite{whitley1999}. Clockwork retains one parent per island, rather than a conventional evolving population within each island. Its combination of broad proposals and local polishing is related to memetic optimization \cite{krasnogor2005}; its fixed repertoire is related to heuristic-selection frameworks \cite{burke2013} and algorithm portfolios \cite{xu2010}. The round-robin scheduler does not learn which operator to select, and the repertoire was not automatically configured. These references establish methodological lineage, not equivalence to those complete algorithms or evidence of improved efficiency.

The archive learner shares the distribution-estimation principle of population-based incremental learning \cite{baluja1994}, but estimates categorical marginals directly from the current elites. Changing perturbation scales and spatial neighborhoods is related to the neighborhood-diversification principle in variable neighborhood search \cite{mladenovic1997}, without implementing its canonical systematic descent and shaking scheme. The exponential uphill-acceptance rule follows the statistical-mechanical motivation of simulated annealing \cite{kirkpatrick1983}; its fixed temperature is stated explicitly below. Together, these connections explain the human-designed search knowledge behind the agents. Their value for this mechanics problem must still be measured through controlled operator, archive, and scheduling ablations.

\section{Design problem and physical evaluation}
\subsection{Categorical geometry and objective}
The supplied structure comprises $30\times30$ cells, each represented by $50\times50$ pixels. The resulting full domain contains $1500\times1500$ bilinear quadrilateral elements. The design is parameterized by a $26\times13$ half-domain label matrix with 168 active categorical entries. Each active entry selects one of 15 permitted candidates from the supplied per-cell library. Fixed and void regions follow the supplied indicator arrays; the half-domain construction is mirrored to obtain the full geometry. Thus, ``random start'' means a random assignment within an existing candidate library, not synthesis of new unit-cell shapes. Formally,
\begin{equation}
 \min_{\mathbf{x}\in\{0,\ldots,14\}^{168}} L(\mathbf{x}),\qquad
 L(\mathbf{x})=\frac{\|P[\mathbf{u}(\mathbf{x})-\mathbf{u}^{\rm ref}]\|_2}
 {\|P\mathbf{u}^{\rm ref}\|_2}.
 \label{eq:objective}
\end{equation}
Here $P$ selects displacement components using the supplied exterior observation mask, $\mathbf{u}^{\rm ref}$ is the supplied reference displacement field without the void, and $\mathbf{u}$ is the computed displacement. All objective nodes must be incident to solid material. An unresolved solve is recorded as an unresolved query and cannot improve the incumbent; it is not assigned an artificial finite displacement loss.

\subsection{Full-resolution equilibrium and void condensation}
All evaluations use the original pixel resolution. The linear plane-stress model uses Poisson's ratio $\nu=0.35$. Opposed horizontal edge forces compress the domain; endpoint nodal loads receive half weights. Horizontal displacement is constrained on the vertical centerline, and vertical displacement at the left, center, and right midpoints. Symmetry reduces the full 4,506,002 displacement degrees of freedom to a half-domain system, after which void-only nodes are eliminated. The source centerline fixes horizontal displacement; midpoint vertical supports and the original loading are retained. Horizontal displacement is reconstructed antisymmetrically and vertical displacement symmetrically. The source element stiffness convention is $K_e=E_0h[E_{\min}+\rho_e(E_0-E_{\min})]\widehat K_e(\nu)$, where $\rho_e\in\{0,1\}$, $E_0=7\times10^6$, $E_{\min}=10^{-9}$, and $h=0.01$. The load parameter is $F_A=3.125\times10^8$. The finite element simulation is solved by conjugate gradients with smoothed-aggregation algebraic multigrid, using SciPy and PyAMG \cite{virtanen2020,bell2023}. Search evaluations use residual tolerance $10^{-8}$ and can reuse solver state. A score is accepted only when the implemented true-residual convergence checks pass. Saved full-domain fields contain zero placeholders at eliminated void-only nodes; these values must not be interpreted as recovered void displacement.

\subsection{Reference and endpoint verification}
Every initial and endpoint design is independently assembled and solved at relative residual tolerance $10^{-10}$. Geometry checks certify supported rigid motions and positive definiteness of the source-Q4 solid system. When an edge-connectivity sufficient condition is inconclusive, the validator can use an exact rational coupled-body rigidity certificate; the criterion is not relaxed merely because a numerical solve converges. For the swarm endpoints reported here, strict residual and geometry certificates are available. A fresh loose-versus-strict repeat and finite-void perturbation diagnostic specifically for the best swarm design are not available in the retained record; no such diagnostic is attributed to it. The zero-void approximation therefore retains its stated scope and limitations.

\section{Clockwork architecture and protocol}
\begin{figure}[!ht]
\centering\includegraphics[width=\linewidth]{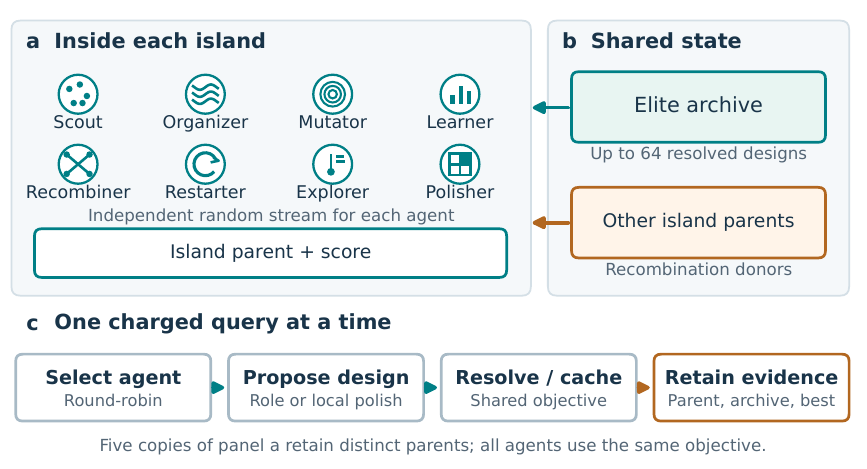}
\caption{Implementation of the swarm. (a) Eight icon-coded proposal roles share an island parent and score, with a separate random stream per agent. This template is repeated in five islands. (b) Archive reads (teal) and cross-island parent reads (orange) provide distinct information channels. (c) A serial scheduler generates, resolves, and retains proposals. Role mechanisms and retention rules are specified in the text and Table~\ref{tab:agent_roles}; global best retention is separate from exploratory parent acceptance.}
\label{fig:clockwork}
\end{figure}

Figure~\ref{fig:clockwork} resolves the conceptual overview in Figure~\ref{fig:concept} into a representative island, its shared information channels, and the serial query cycle. The following subsections specify agent responsibilities, information exchange, scheduling, and retention.

\subsection{Division of concerns: agents within islands}
Clockwork is a swarm of forty stateful numerical search agents organized into five cooperating islands (Figure~\ref{fig:clockwork}). Each island contains eight operators with distinct proposal responsibilities. Each agent owns its random-number stream, proposal counter, and evaluation statistics. Eight agents within an island act on a common parent assignment and physical score, whereas the five islands retain separate parents. This organization separates proposal diversity, local state retention, and shared physical evidence. Here, autonomy means generating a candidate from a specified policy and state; it does not imply language-model reasoning, inter-agent dialogue, or a separate computing node.

For reproducible identification, agent $A_{8(i-1)+r}$ belongs to island $I_i$ and role $R_r$, with $i=1,\ldots,5$ and $r=1,\ldots,8$. These manuscript identifiers are one-based; the event log uses zero-based agent indices 0--39. Thus $I_1$ comprises $A_1$--$A_8$, $I_2$ comprises $A_9$--$A_{16}$, and so forth. Table~\ref{tab:agent_roles} specifies the responsibility of all forty agents through this repeated role map. A candidate always assigns all 168 active cells: agents do not own disjoint geometric subdomains or solve separate stress and displacement objectives.

\begingroup\small
\begin{longtable}{p{.20\linewidth}p{.23\linewidth}p{.47\linewidth}}
\caption{Agent responsibilities. Each row is instantiated once in each of the five islands; listed agent identifiers therefore cover all forty agents. Descriptions apply to specialty actions; the interleaved local-polishing schedule is described below.}\label{tab:agent_roles}\\
\toprule Role and agents & Search concern and input & Responsibility and proposal mechanism\\\midrule
\endfirsthead
\multicolumn{3}{l}{\textit{Table~\ref{tab:agent_roles} continued}}\\\toprule Role and agents & Search concern and input & Responsibility and proposal mechanism\\\midrule
\endhead
\bottomrule\endfoot
$R_1$: Global scout\newline $A_{1,9,17,25,33}$ & Unbiased exploration; label alphabet & Independently samples every active cell uniformly from the 15 labels, allowing a proposal far from the current island parent.\\[5pt]
$R_2$: Spatial organizer\newline $A_{2,10,18,26,34}$ & Spatial coherence; active-cell coordinates & Tests homogeneous assignments first, covering all 15 labels across the islands' first three specialty rounds. Subsequently cycles through Voronoi, stripe, and block patterns.\\[5pt]
$R_3$: Mutation specialist\newline $A_{3,11,19,27,35}$ & Perturbation scale; island parent & Selects a fraction from $\{1/n,0.03,0.1,0.3,0.7,1\}$ and changes that many entries, with $n=168$ and rounding up. Every selected entry receives a different label.\\[5pt]
$R_4$: Archive learner\newline $A_{4,12,20,28,36}$ & Reuse of good categorical choices; up to 24 elites & Learns per-cell label frequencies from resolved archived assignments, mixed with 35\% uniform sampling. This is a categorical sampling model, not a displacement surrogate.\\[5pt]
$R_5$: Recombiner\newline $A_{5,13,21,29,37}$ & Exchange of design content; island parents and archive & Selects a parent from the five islands, with a 50\% chance of replacing that donor by an archive member. Uses a random 0.2--0.8 mixing probability and independently resamples each entry with probability 0.1.\\[5pt]
$R_6$: Diversity restarter\newline $A_{6,14,22,30,38}$ & Escape from a parent trajectory; island parent & Draws a new uniform assignment and retains each parent entry with probability 0.2. Worse resolved proposals can be accepted, preserving exploratory trajectories.\\[5pt]
$R_7$: Neighborhood explorer\newline $A_{7,15,23,31,39}$ & Spatially localized collective changes; coordinates and parent & Chooses a random active-cell center and changes the nearest $\lceil n\alpha\rceil$ entries, with $\alpha\sim U(0.1,0.8)$. Temperature-based acceptance permits uphill parent moves.\\[5pt]
$R_8$: Incumbent polisher\newline $A_{8,16,24,32,40}$ & Fine refinement; best archived assignment & Changes one entry of the global incumbent using an island-offset coordinate sequence and cyclic alternative labels. It can refine another island's discovery through the shared archive.\\
\end{longtable}
\endgroup

The seven broad specialties ($R_1$--$R_7$) each interleave one specialty proposal with two sparse local-polishing proposals that change one to three entries of their own island parent. The dedicated $R_8$ agents instead probe the global incumbent. Consequently, a declared specialty identifies the agent's responsibility, but not every action it executes. The event record retains both fields. This distinction prevents an improvement made by a restarter's local-polish action from being misreported as evidence that a broad restart was effective.

\subsection{Responsibilities of individual agents}
The role map in Table~\ref{tab:agent_roles} specifies the implemented actions. The following explanations connect each duty to the search principles reviewed above. They describe design rationales, not measured causal contributions from role-ablation experiments.

\paragraph{\agenticon{1}$R_1$: Global scout---generate alternatives without parent bias.}
The scout samples the entire categorical assignment independently of the current island parent and shared archive, exploring regions of the design space that incremental refinement may not reach. Each evaluated candidate receives a physical score and may become a new island parent or enter the elite archive. Unlike the spatial organizer, the scout imposes no spatial pattern on its proposals.

\paragraph{\agenticon{2}$R_2$: Spatial organizer---propose coherent arrangements.}
The organizer uses active-cell coordinates to generate homogeneous, Voronoi, stripe, and block assignments. It introduces spatial correlation into the categorical search, rather than assuming independent cell choices are sufficient. The initial homogeneous tests span all fifteen labels across the five islands. Later patterned proposals can seed the same archive used by learners and recombiners. Spatial coherence is a proposal prior; it does not guarantee compatibility, connectivity, or a low physical loss.

\paragraph{\agenticon{3}$R_3$: Mutation specialist---bridge small and large changes.}
The mutation agent begins from its island parent and varies the fraction of altered entries. Its duty is to test whether improvement requires a small adjustment or a substantial rearrangement while preserving the remainder of an existing design. Selected entries receive different labels. Unlike the neighborhood explorer, selection need not be spatially contiguous. The resulting candidates connect local exploitation and broad exploration without changing the physical objective.

\paragraph{\agenticon{4}$R_4$: Archive learner---reuse successful categorical choices.}
The learner reads up to twenty-four elite assignments and estimates a label-frequency distribution separately at each active cell, following the distribution-estimation principle discussed in \cite{baluja1994}. Its duty is to convert shared evaluated designs into a new sampling policy. Mixing empirical frequencies with uniform sampling keeps all labels accessible. Because this model uses marginal frequencies, it does not explicitly preserve correlations between cells or predict displacement. Its proposals therefore require the same geometry and equilibrium evaluation as all other candidates.

\paragraph{\agenticon{5}$R_5$: Recombiner---transfer design content between trajectories.}
The recombiner mixes its own island parent with a donor drawn from the island parents or the shared archive, then applies categorical resampling. Its duty is direct exchange of design content across search histories. This differs from the learner's statistical reuse: recombination copies entries from complete assignments. The donor can be the same island, so not every recombination is a cross-island exchange. A combined assignment is evaluated afresh because good parent scores do not imply a good combined mechanical response.

\paragraph{\agenticon{6}$R_6$: Diversity restarter---renew a parent trajectory.}
The restarter generates a mostly new assignment while probabilistically retaining entries from its own parent. Its duty is to provide an escape mechanism when a parent trajectory becomes restrictive. The acceptance rule can retain a worse resolved parent, whereas the global archive protects the best known design. This architectural separation permits exploration without erasing the incumbent. The duty is to create opportunities for renewed search, not to guarantee or maximize a measured population-diversity metric.

\paragraph{\agenticon{7}$R_7$: Neighborhood explorer---change spatial groups together.}
The neighborhood explorer chooses an active-cell center and perturbs a group of nearby cells. It tests coordinated spatial changes that a one-cell update cannot represent. Unlike the spatial organizer, it preserves the parent outside the selected neighborhood; unlike unstructured mutation, it selects entries by spatial distance. Temperature-based acceptance permits uphill moves \cite{kirkpatrick1983}. The name ``annealed neighborhood'' in the source denotes this acceptance mechanism; the implemented temperature is fixed from the initial loss rather than cooled over time.

\paragraph{\agenticon{8}$R_8$: Incumbent polisher---refine the shared best design.}
The polisher reads the global incumbent from the archive and probes one coordinate at a time using an island-offset sequence. It performs focused refinement of the swarm's best available assignment, including one discovered by another island. This separation of broad variation and fine refinement follows the hybrid-search rationale of memetic methods \cite{krasnogor2005}. The other seven roles also perform interleaved sparse local polishing, but those actions start from their own island parent. Thus local parent refinement and shared-incumbent refinement remain distinct.

\subsection{What distinguishes the five islands}
The five islands maintain separate search trajectories while sharing the same objective, geometry constraints, operator repertoire, and acceptance rules. At initialization they receive the same run-specific random assignment; independent agent random streams subsequently produce different proposals and parent histories. Deterministic island offsets also distinguish the initial homogeneous-label allocation and the dedicated polishing coordinate sequence. The homogeneous-label permutation is generated once per run, so this offset distributes the 15 homogeneous tests without assigning a permanently preferred label to an island.

Each island is responsible for retaining one exploratory parent while its eight agents explore and refine alternatives. Island $I_1$ owns $A_1$--$A_8$, $I_2$ owns $A_9$--$A_{16}$, $I_3$ owns $A_{17}$--$A_{24}$, $I_4$ owns $A_{25}$--$A_{32}$, and $I_5$ owns $A_{33}$--$A_{40}$. These identifiers are stable throughout a run; each island develops a distinct search history through its independent proposals and parent updates. Because equilibrium couples all cells, all islands address the complete assembly problem.

\subsection{How the islands work together}
Coordination occurs through explicit design-state exchange rather than all-to-all messaging. Within an island, an accepted candidate becomes the parent seen by its other seven agents on later turns. Across islands, three channels transfer information: (i) any resolved proposal can enter the shared elite archive, (ii) recombination can read another island parent or an archived assignment, and (iii) incumbent polishers start from the best archived design regardless of which island discovered it. Archive learners additionally convert elite assignments into per-cell categorical probabilities. Thus the swarm shares both complete designs and a statistical summary of successful design choices.

A typical cooperation cycle is as follows. A spatial organizer proposes a coherent assignment; the physical evaluator returns its displacement loss; a resolved good assignment enters the shared archive. An archive learner can subsequently reuse its label frequencies, a recombiner can mix its entries with another parent, and an incumbent polisher can refine the resulting best design one coordinate at a time. This describes an available information path, not a reconstructed causal history of a particular endpoint. The current event log records agent actions and losses, but not every donor-parent identity or archive-read lineage.

The shared archive does not force all island parents to become the same design. Parent acceptance is local to the selected island and may permit a worse parent for designated exploratory actions, while global best-design retention is separate. This separation allows an island to explore without discarding the swarm's best known assignment. All agents receive physical objective feedback through the same evaluator; geometry certification, equilibrium solution, caching, and final strict validation are infrastructure services, not additional agents or island-specific responsibilities. No displacement surrogate, supplied optimized assignment, or target-based stopping rule guides proposals.

\subsection{State, proposal scheduling, and retention}
At step $t$, island $i\in\{1,\ldots,5\}$ holds a parent $\mathbf p_i$ and score $s_i$. Each of its eight operators has an independent random stream. A deterministic round-robin scheduler selects one of forty agents. The order is $I_1:R_1,\ldots,R_8$, then $I_2:R_1,\ldots,R_8$, through $I_5:R_8$, repeated thereafter. Repeated within-run assignments are skipped without advancing the agent turn, although their proposal attempts advance that agent's proposal counter. A charged distinct assignment advances the turn. For the seven broad specialties, one specialty proposal alternates with two sparse local-polish proposals; the dedicated polishing specialty follows its own one-coordinate sequence. This schedule is independent of the final budget.

The archive stores at most 64 resolved queried assignments, sorted by loss. Categorical learning uses up to 24 elites: for cell $j$ and label $k$, its sampling distribution is
\begin{equation}
 q_{jk}=0.65\,\frac{1}{m}\sum_{\ell=1}^{m}
 \mathbf1[x_j^{(\ell)}=k]+\frac{0.35}{15},\qquad m\le24.
\end{equation}

The first resolved agent proposal for each island initializes its parent even when worse than the shared initial assignment. Subsequently, a resolved improving proposal replaces its island parent. Selected exploratory roles may accept a worse parent: uniform sampling and diversity reset use probability 0.35, whereas correlated fields, recombination, and annealed neighborhoods use $\exp[\min(0,(s_i-L)/T)]$, with $T=\max(10^{-8},0.15|L_0|)$. These choices preserve distinct parent trajectories. The reported global incumbent is the resolved assignment with the lowest displacement loss, so its loss remains nonincreasing throughout the search. Unresolved proposals cannot enter the elite archive or replace an island parent. The frozen source defines the complete proposal distributions and duplicate handling; the present equations describe the implemented mechanisms rather than a convergence guarantee.

\subsection{Random starts and query accounting}
Seeds 0, 1, and 2 define three random geometry-certified starting assignments. Initialization does not screen physical loss. Each run maintains its own optimizer state, and no supplied optimized assignment guides proposals. The budget includes the starting design and every distinct attempted assignment. Shared-cache hits and unresolved outcomes consume charged queries; repeated within-run assignments are skipped. The physical cache was maintained by the broader recorded campaign, so cache hits are charged even when the underlying solve was already available. This reporting subset is not a newly executed isolated-cache experiment.

Physical evaluations execute serially. Forty logical agents therefore do not imply forty concurrent solves or a measured wall-clock speedup.

\subsection{Retrospective endpoint recovery}
The original campaign progressed through 301 and 601 queries and was extended toward 1,201. The investigator selected 961 after observing progress. The incumbent is reconstructed exclusively from events 1--961 of each swarm run, with label hashes and the original 301- and 601-event prefix hashes checked against preserved manifests. Later events cannot affect the reported designs. This recovery requires no new optimization. The 961-query endpoint was selected retrospectively from a longer campaign and is reported as an exploratory result.

The three swarm runs account for $3\times961=2,883$ charged queries, including 1,080 after their 601-query checkpoints. Counts here refer only to the swarm subset of the archived campaign.

\section{Field reconstruction and post-processing}
The retained endpoint labels define fixed geometries for three additional strict equilibrium solves, outside the optimization budgets. Their losses are compared with the archived strict results before post-processing. The full nodal displacement is reconstructed using antisymmetric horizontal and symmetric vertical reflection. A Q4 interpolation gives the element-center displacement
\begin{equation}
 \mathbf u_e^c=\sum_{a=1}^{4}N_a(0,0)\mathbf u_a
 =\tfrac14\sum_{a=1}^{4}\mathbf u_a,\qquad
 |\mathbf u_e^c|=\sqrt{(u_{x,e}^c)^2+(u_{y,e}^c)^2}.
\end{equation}
Nodal values at eliminated void-only degrees of freedom are zero placeholders, not recovered void displacements. All displayed displacement and stress fields therefore mask void elements. Displacements retain the original model units, without a conversion to millimeters.

For a solid element, the normalized stress vector is obtained from
\begin{equation}
 \widetilde{\boldsymbol\sigma}_e
 =\frac{D(\nu)B_e(0,0)\mathbf u_e}{t_0},\qquad
 t_0=\frac{F_A}{1500 E_0^2h},\qquad
 D(\nu)=\frac{1}{1-\nu^2}
 \begin{bmatrix}1&\nu&0\\\nu&1&0\\0&0&(1-\nu)/2\end{bmatrix}.
 \label{eq:stress}
\end{equation}
Unit pixel length and the original Q4 node ordering are retained in $B_e$. This normalization is the ratio to applied nominal traction under the same legacy scaling. Physical $x$ points right and $y$ points up, whereas stored image rows increase downward. Negative normal stress denotes compression.

The equivalent and in-plane principal stresses are
\begin{align}
 \widetilde\sigma_{\rm vm}
 &=\sqrt{\widetilde\sigma_{xx}^2-\widetilde\sigma_{xx}\widetilde\sigma_{yy}
 +\widetilde\sigma_{yy}^2+3\widetilde\tau_{xy}^2},\\
 \widetilde\sigma_{1,2}
 &=\tfrac12(\widetilde\sigma_{xx}+\widetilde\sigma_{yy})
 \mathbin{\pm}\sqrt{\tfrac14(\widetilde\sigma_{xx}-\widetilde\sigma_{yy})^2+
 \widetilde\tau_{xy}^2}.
\end{align}
The out-of-plane normal and transverse shear stresses vanish in the plane-stress model. The two displayed principal values are the in-plane eigenvalues, not an ordering of all three principal stresses in three dimensions. Center stress equals the average of the four Gauss-point stress vectors for this constant-material element; center von Mises stress is not the average of Gauss-point von Mises values. These numerical verification checks establish consistency within the discrete model but do not demonstrate continuum convergence or physical calibration.

\section{Results}
\subsection{Independent endpoint validation}
Table~\ref{tab:results} reports strict displacement losses for the three swarm runs. The arithmetic endpoint mean is 0.125240554 and the sample standard deviation is 0.021416579. These statistics summarize three independent search runs on a single design problem and do not establish performance consistency across different problem configurations.
\begin{table}[!htb]\centering
\caption{Swarm displacement losses from fresh $10^{-10}$ solves. Initial and final assignments pass their geometry and residual checks.}\label{tab:results}
\begin{tabular}{rrr}\toprule Seed & Random-start loss & 961-query loss\\\midrule
0 & 0.649174685 & 0.100589562\\
1 & 0.435813867 & 0.139273770\\
2 & 0.383127887 & 0.135858330\\\midrule
Mean & 0.489372147 & 0.125240554\\\bottomrule\end{tabular}\end{table}

\begin{figure}[!htb]\centering\includegraphics[width=\linewidth]{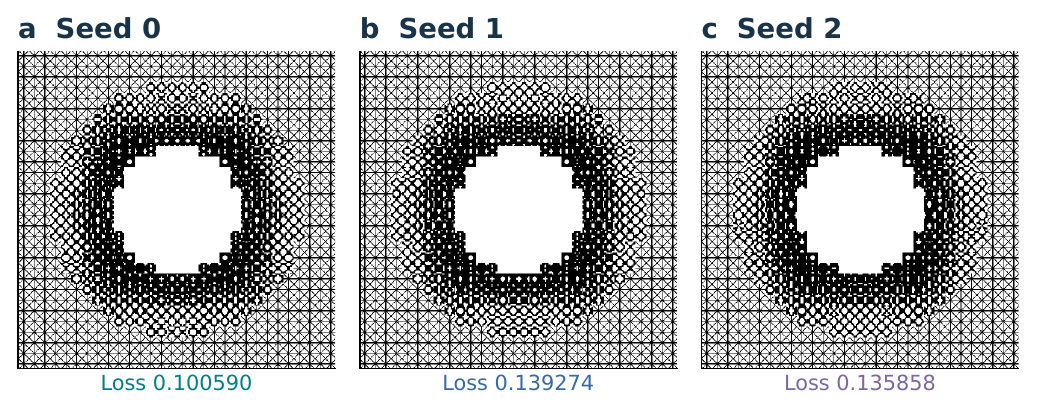}
\caption{Recovered swarm geometries at exactly 961 queries: (a) seed 0, (b) seed 1, and (c) seed 2. Black represents solid pixels and white void. Images use the verified full-resolution label assemblies, rather than generated illustrations. Labels beneath each panel give strict displacement loss.}\label{fig:designs}\end{figure}

The best design is swarm seed 0, with strict loss $L_{\min}=0.100589561695$. Relative to the comparison threshold $\tau=0.1102$, its absolute margin below threshold is 0.009610438305, its ratio $L_{\min}/\tau$ is 0.9127909410, and its loss is 8.72\% below threshold. Only seed 0 reaches the threshold; it is a comparison value rather than an optimality certificate. Figure~\ref{fig:designs} shows all three recovered endpoint geometries.

\subsection{Search trajectories and budget dependence}
\begin{figure}[p]\centering\includegraphics[width=\linewidth,height=.78\textheight,keepaspectratio]{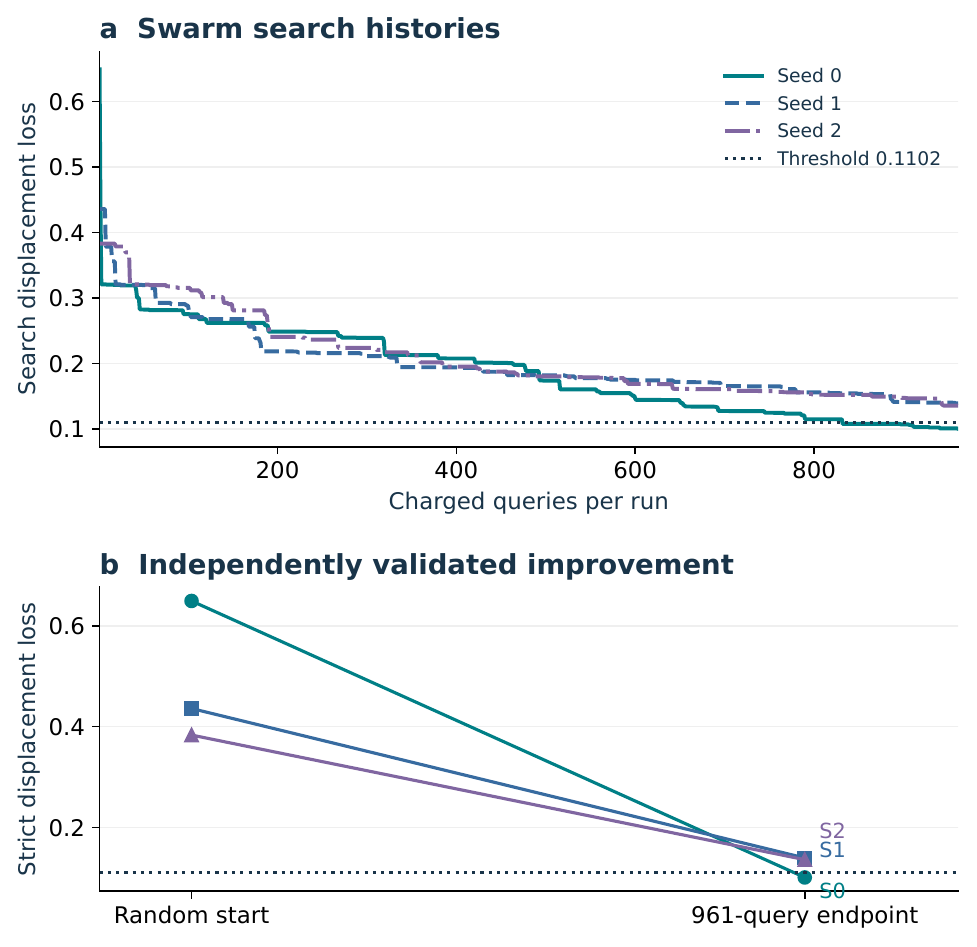}
\caption{Swarm performance. (a) Incumbent search-tolerance loss over the first 961 charged queries, with color and line style identifying seed. (b) Independent strict losses of each random start and final assignment. The dotted line marks 0.1102. No later queries contribute, and no confidence interval is inferred from three seeds.}\label{fig:curves}\end{figure}

Figure~\ref{fig:curves}(a) reports the three search histories at $10^{-8}$ tolerance; panel (b) uses independent strict start and endpoint losses. The mean strict loss decreases from 0.162784021 at 601 queries to 0.125240554 at 961, a 23.06\% reduction. At search tolerance, the arithmetic means are 0.224935, 0.162784, 0.138348, and 0.132173 at 301, 601, 841, and 901 queries, respectively. Seed 0 first records a threshold-reaching incumbent at query 832. That first-crossing assignment was not separately strict-validated; the reported strict result belongs to its 961-query incumbent.

\subsection{Accounting and numerical consistency}
The swarm subset contains 205 unresolved events and 32 cache-hit events among its 2,883 charged queries. These classifications can overlap. Per-seed unresolved counts are 74, 78, and 53; cache-hit counts are 6, 13, and 13. Duplicate-proposal counts were unavailable at the 961-query cutoff, and optimizer wall time at exactly query 961 cannot be reconstructed from the event record.

All three swarm endpoints and their three starts pass independent strict solves. Endpoint true relative residuals are $9.48\times10^{-11}$, $8.89\times10^{-11}$, and $7.86\times10^{-11}$. The maximum absolute search-to-strict loss shift among the swarm endpoints is $1.63\times10^{-11}$. This compares the recorded search loss with a fresh strict solve; it is not an additional fresh loose/strict experiment. Exact coupled-body rigidity certificates support seeds 0 and 1 when the simpler edge-connectivity gate is inconclusive; seed 2 passes the sufficient edge-connectivity certificate. These certify the constrained source-discrete solid system, not manufacturability or finite-void approximation error.

\subsection{Independent field reconstruction and verification}
The additional field-recovery solves reproduce all three archived strict losses exactly at the recorded precision. Their relative residuals remain below $10^{-10}$, and their label hashes match the frozen endpoint records. Table~\ref{tab:verification} separates algebraic convergence, geometry certification, and post-processing checks. 

\begin{table}[htbp]\centering\small
\caption{Scope of numerical verification for the fixed swarm endpoints and stress-recovery implementation.}\label{tab:verification}
\begin{tabular}{p{.48\linewidth}p{.42\linewidth}}\toprule
Check & Recorded evidence\\\midrule
Endpoint residuals, seeds 0--2 & $(9.48,8.89,7.86)\times10^{-11}$\\
Independent field-solve loss differences & Zero at recorded double precision\\
Search-to-strict loss shift, maximum & $1.63\times10^{-11}$\\
Q4 stiffness quadrature, maximum entry difference & $1.11\times10^{-16}$\\
Affine stress patch, maximum discrepancy & $2.17\times10^{-19}$\\
Source-discrete rigidity & Exact coupled-body checks for seeds 0--1; sufficient edge-connectivity check for seed 2\\\bottomrule
\end{tabular}\end{table}

\subsection{Displacement fields}
\begin{figure}[p]\centering\includegraphics[width=\linewidth,height=.73\textheight,keepaspectratio]{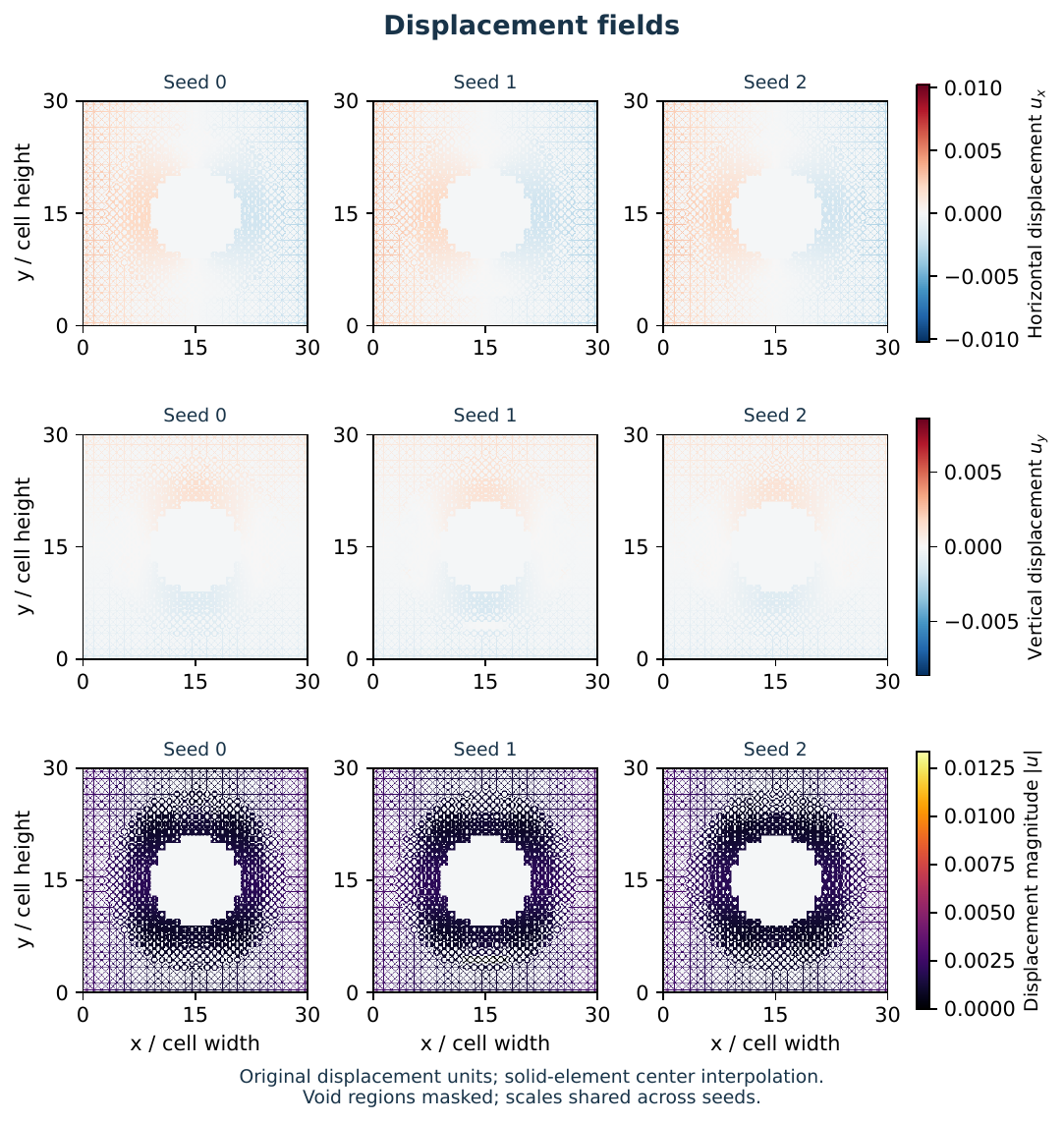}
\caption{Displacements of the three fixed 961-query swarm endpoints. Columns correspond to seeds 0, 1, and 2; rows show $u_x$, $u_y$, and $|\mathbf u|$. Values are solid-element center interpolations of freshly recovered strict nodal fields. Void elements are masked.}\label{fig:displacements}\end{figure}

Figure~\ref{fig:displacements} displays horizontal and vertical displacement and their magnitude on the solid geometry. The horizontal field follows the opposed compressive loading; the vertical field reflects the centerline and midpoint constraints together with the heterogeneous assembly. Their visual similarity across seeds is consistent with a common loading and fixed exterior microstructure, but visual similarity is not the optimized norm. Equation~\ref{eq:objective} uses only the prescribed exterior observation coordinates, whereas the maps show the entire solid region.

Maximum solid element-center displacement magnitudes are 0.013326, 0.013285, and 0.013348 in the source model units. Similar full-domain maxima accompany substantially different exterior losses. Consequently, maximum displacement is not a surrogate for cloak quality, and the smallest exterior norm error does not imply the smallest maximum displacement. The global magnitude field and the objective measure different aspects of the mechanical response.

\subsection{Stress components and local load transfer}
\begin{figure}[p]\centering\includegraphics[width=\linewidth,height=.73\textheight,keepaspectratio]{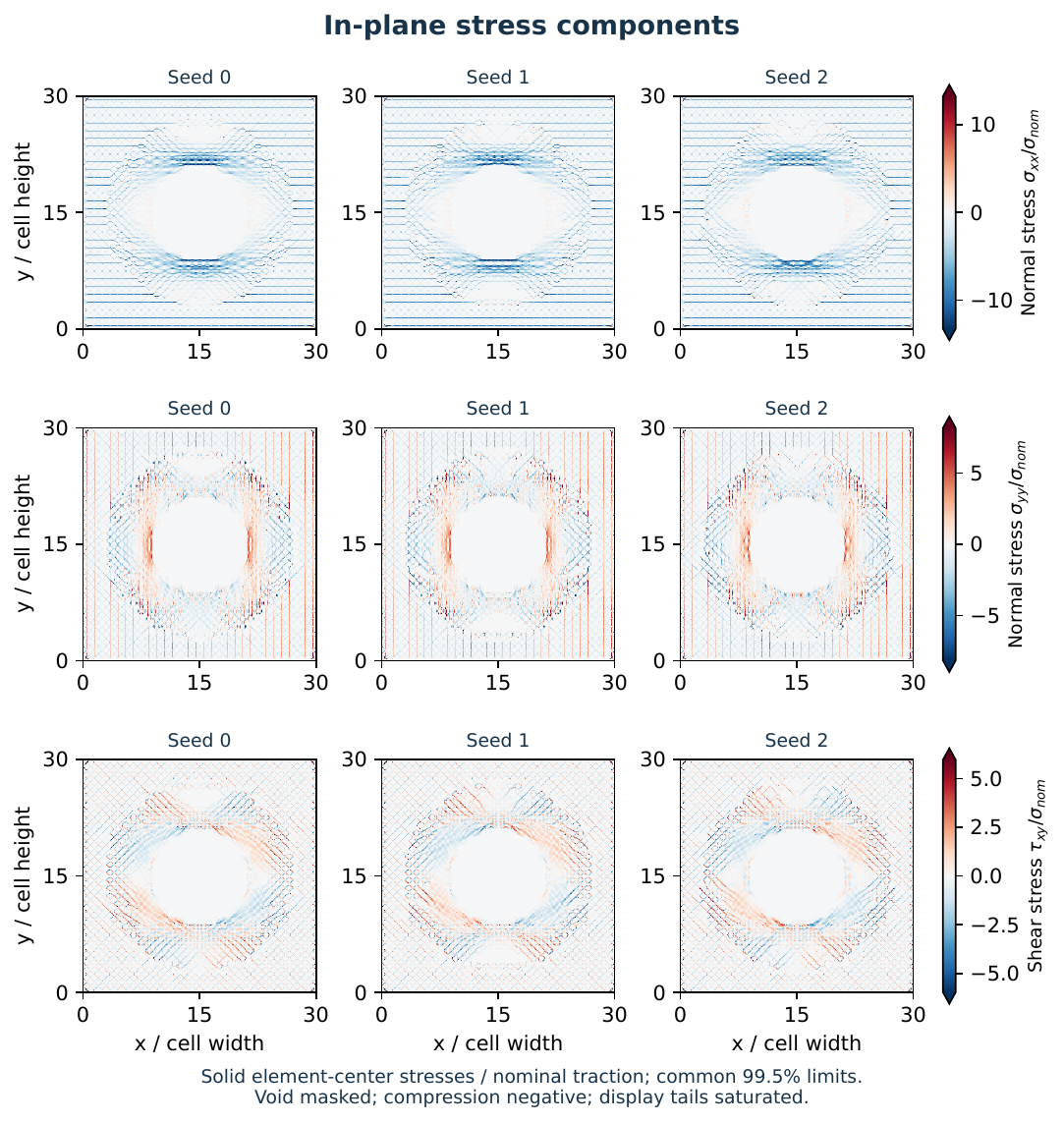}
\caption{Normalized solid-element center stress components of the three fixed swarm endpoints. Columns identify seeds 0--2 and rows show $\sigma_{xx}$, $\sigma_{yy}$, and $\tau_{xy}$ divided by nominal traction. Compression is negative. Each row has a common symmetric display scale; colorbar extensions indicate saturated tails. Void elements are masked.}\label{fig:stress}\end{figure}

Figure~\ref{fig:stress} resolves the two normal stresses and in-plane shear stress. Stronger horizontal compression occurs in bands above and below the cavity, while transverse and shear components redistribute around the cavity and heterogeneous cell connections. 

\begin{figure}[p]\centering\includegraphics[width=\linewidth,height=.73\textheight,keepaspectratio]{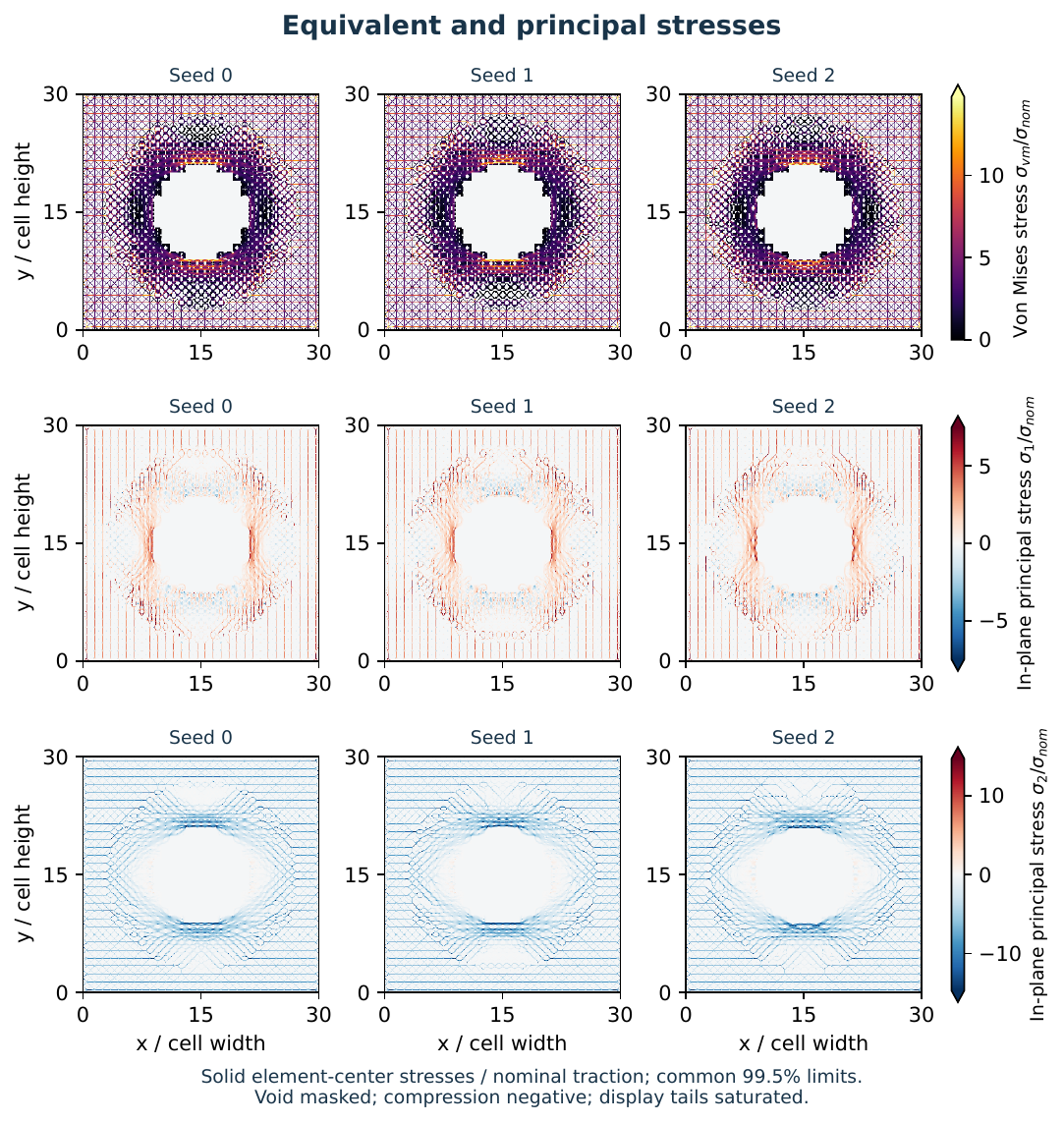}
\caption{Equivalent and principal stresses for the same fixed designs. Rows show normalized von Mises stress and the maximum and minimum in-plane principal stresses. The out-of-plane principal value is zero in this plane-stress model. Scales are shared across seeds separately for each row; percentile display clipping and void masking follow Figure~\ref{fig:stress}. Unclipped arrays and extrema are retained for quantitative use.}\label{fig:principal}\end{figure}

Figure~\ref{fig:principal} reports von Mises and in-plane principal stresses. The 99.5th percentiles of solid-element von Mises stress, normalized by nominal traction, are 14.277, 14.728, and 14.588 for seeds 0--2. The corresponding unclipped maxima are approximately 106.663 in all three cases. This large separation between bulk upper-tail values and extrema makes peak-based conclusions particularly sensitive to pixel corners, thin connections, and constrained or loaded nodes. 

Stress display limits use a common pooled absolute 99.5th percentile for each component, estimated from deterministic every-fourth solid-element samples. Both tails of signed fields are saturated beyond the shared limits; von Mises is clipped only above its positive limit. Exact per-case distributions and unclipped values are preserved in the numerical field files. The percentile values reported in the text use all solid elements and are distinct from the pooled display limits. 

\subsection{Observed participation of the cooperating islands}
At the 961-query endpoint, each run contains one initial evaluation and 960 agent evaluations. Every agent therefore contributes exactly 24 charged distinct proposals, and every island contributes 192, irrespective of its achieved loss. Table~\ref{tab:islands} reports the number of improvements in the best-so-far displacement loss attributed to proposals from each island. All five islands contribute such improvements in each run; the relative contributions differ across starts. These counts use recorded search evaluations; intermediate improvements were not individually recomputed at the stricter tolerance.

\begin{table}[htbp]\centering\small
\caption{Realized island participation at 961 queries. Each island has the same eight-role team and receives 192 charged proposals per seed. Entries count global-incumbent improvements attributed to that island, excluding initialization.}\label{tab:islands}
\begin{tabular}{llrrr}\toprule
Island & Agent identifiers & Seed 0 & Seed 1 & Seed 2\\\midrule
$I_1$ & $A_{1}$--$A_{8}$ & 14 & 32 & 11\\
$I_2$ & $A_{9}$--$A_{16}$ & 23 & 21 & 7\\
$I_3$ & $A_{17}$--$A_{24}$ & 11 & 19 & 19\\
$I_4$ & $A_{25}$--$A_{32}$ & 23 & 8 & 28\\
$I_5$ & $A_{33}$--$A_{40}$ & 22 & 12 & 23\\
\bottomrule\end{tabular}
\end{table}

The final retained search design in seed 0 is proposed at query 961 by $A_{40}$ in $I_5$, acting as the incumbent polisher. In seed 1, $A_{38}$ in $I_5$ proposes the final retained design at query 959 while performing interleaved island-local polishing, despite its declared diversity-reset specialty. In seed 2, $A_{24}$ in $I_3$ proposes the retained design at query 945 as an incumbent polisher. These observations illustrate why agent identity, declared responsibility, and executed action must be distinguished. They do not show that the final proposing agent independently created the design: its input state can incorporate earlier work shared through the archive and island parents.

An improvement count is not an efficiency ranking. A broad proposal can supply useful diversity without immediately setting a record, while a polishing proposal can exploit a discovery made elsewhere. Counts depend on sequential ordering, shared state, and the common starting design. The present logs do not isolate counterfactual performance with an island or communication channel removed; controlled ablations remain necessary to quantify the value of cooperation.

\section{Discussion}
\subsection{What the portfolio demonstrates}
Clockwork finds a strictly verified sub-threshold displacement design from a random assignment within the supplied candidate sets. Its continued mean improvement between 601 and 961 queries shows that the recorded search had not exhausted its useful refinement at the earlier checkpoint. Separate island parents and shared elite information permit different trajectories, while the global archive protects the best queried design from exploratory parent updates.

This is evidence of feasibility on one demanding discrete assembly problem, not an attribution of gain to agent count or communication. No role ablation, communication ablation, agent-count sweep, or controlled baseline is included in this swarm-only manuscript. Accepted proposals are conditioned on the evolving archive and are not independent treatment effects. Comparative efficiency, robustness across problems, and parallel speedup therefore remain untested.

\subsection{Mechanical interpretation of the recovered fields}
The field plots add physical interpretation to a scalar search record. Displacement maps show that comparable full-domain response amplitudes can coexist with different masked exterior errors. Stress maps identify localized load-transfer bands and strong upper tails that are not controlled by a displacement-only objective. The best displacement design should consequently not be described as a stress-optimal, strength-optimal, or failure-resistant cloak.

\subsection{Implications for further method development}
A substantive next test would prescribe additional random starts and multiple load or cavity configurations before observing their outcomes, retaining the same accounting and validation rules. Operator and archive ablations would isolate which mechanisms justify the portfolio's complexity. Finite-contrast and resolution studies are needed before transferring local field conclusions to a physical continuum or manufacturing process.

\section{Limitations and reproducibility}
The evidence covers one geometry, one loading/observation configuration, three starts, and one portfolio configuration. Selecting the endpoint after observing progress adds selection risk. Geometry-conditioned initialization, shared caching, and unresolved-query accounting are part of the recorded protocol. A supplied unit-cell library and reference field remain essential inputs even though no supplied optimized assignment guides search.

\section{Conclusion}
Clockwork's forty numerical specialists, organized into five communicating islands, produce strict endpoint losses of 0.100590, 0.139274, and 0.135858 after exactly 961 charged queries from three random starts. The best loss is 8.72\% below 0.1102, and the mean decreases by 23.06\% from the validated 601-query checkpoint. Independent strict solves and source-discrete rigidity checks support the reported physical evaluations. Fresh field reconstruction reproduces the saved losses, while stiffness quadrature and affine patch checks verify stress recovery. The displacement and stress maps expose distinct response measures and localized stress concentrations that a displacement objective alone does not constrain. Broader benchmarks, controlled ablations, and finite-contrast and mesh studies are needed before comparative efficiency or physical robustness can be established.
\section*{Statements and declarations}
\paragraph{Data and code availability.} The current manuscript, editable source, plotting scripts, recovered endpoint labels, recorded validation, and frozen optimizer/solver sources are available at \url{https://github.com/BaratiLab/Metamaterial-Swarm}, with SHA-256 integrity manifests and a read-only record-verification script. Original input libraries, observation masks, reference fields, and full raw nodal-field caches are not redistributed. The study guide states these reproduction requirements. A versioned archival DOI and input redistribution permissions remain to be finalized by the authors.

\paragraph{Artificial-intelligence assistance.} 
A language-model assistant supported implementation, figure preparation, analysis, and manuscript drafting. The forty search agents studied here are numerical operators, not language models. The authors are responsible for verifying the scientific content.

\paragraph{Acknowledgments.}
The authors thank Dr. Liwei Wang for providing the dataset and simulation code and for discussions.


\begin{thebibliography}{99}
\bibitem{wang2022} L. Wang, J. Boddapati, K. Liu, P. Zhu, C. Daraio, and W. Chen, ``Mechanical cloak via data-driven aperiodic metamaterial design,'' \emph{Proceedings of the National Academy of Sciences}, vol. 119, no. 13, e2122185119, 2022. \href{https://doi.org/10.1073/pnas.2122185119}{doi:10.1073/pnas.2122185119}.
\bibitem{shapedna} L. Wang, Y.-C. Chan, Z. Liu, P. Zhu, and W. Chen, ``Data-driven metamaterial design with Laplace--Beltrami spectrum as shape-DNA,'' \emph{Structural and Multidisciplinary Optimization}, vol. 61, pp. 2613--2628, 2020. \href{https://doi.org/10.1007/s00158-020-02523-5}{doi:10.1007/s00158-020-02523-5}.
\bibitem{wang2020} L. Wang, Y.-C. Chan, F. Ahmed, Z. Liu, P. Zhu, and W. Chen, ``Deep generative modeling for mechanistic-based learning and design of metamaterial systems,'' \emph{Computer Methods in Applied Mechanics and Engineering}, vol. 372, 113377, 2020. \href{https://doi.org/10.1016/j.cma.2020.113377}{doi:10.1016/j.cma.2020.113377}.
\bibitem{metaset} Y.-C. Chan, F. Ahmed, L. Wang, and W. Chen, ``METASET: Exploring shape and property spaces for data-driven metamaterials design,'' \emph{Journal of Mechanical Design}, vol. 143, no. 3, 031707, 2021. \href{https://doi.org/10.1115/1.4048629}{doi:10.1115/1.4048629}.
\bibitem{panetta} J. Panetta, Q. Zhou, L. Malomo, N. Pietroni, P. Cignoni, and D. Zorin, ``Elastic textures for additive fabrication,'' \emph{ACM Transactions on Graphics}, vol. 34, no. 4, 135, 2015. \href{https://doi.org/10.1145/2766937}{doi:10.1145/2766937}.
\bibitem{schumacher} C. Schumacher, B. Bickel, J. Rys, S. Marschner, C. Daraio, and M. Gross, ``Microstructures to control elasticity in 3D printing,'' \emph{ACM Transactions on Graphics}, vol. 34, no. 4, 136, 2015. \href{https://doi.org/10.1145/2766926}{doi:10.1145/2766926}.
\bibitem{xu2010} L. Xu, H. H. Hoos, and K. Leyton-Brown, ``Hydra: Automatically configuring algorithms for portfolio-based selection,'' \emph{Proceedings of the AAAI Conference on Artificial Intelligence}, vol. 24, no. 1, pp. 210--216, 2010. \href{https://doi.org/10.1609/aaai.v24i1.7565}{doi:10.1609/aaai.v24i1.7565}.
\bibitem{bertoldi2017} K. Bertoldi, V. Vitelli, J. Christensen, and M. van Hecke, ``Flexible mechanical metamaterials,'' \emph{Nature Reviews Materials}, vol. 2, 17066, 2017. \href{https://doi.org/10.1038/natrevmats.2017.66}{doi:10.1038/natrevmats.2017.66}.
\bibitem{bendsoe1988} M. P. Bends\o e and N. Kikuchi, ``Generating optimal topologies in structural design using a homogenization method,'' \emph{Computer Methods in Applied Mechanics and Engineering}, vol. 71, no. 2, pp. 197--224, 1988. \href{https://doi.org/10.1016/0045-7825(88)90086-2}{doi:10.1016/0045-7825(88)90086-2}.
\bibitem{sigmund1994} O. Sigmund, ``Materials with prescribed constitutive parameters: An inverse homogenization problem,'' \emph{International Journal of Solids and Structures}, vol. 31, no. 17, pp. 2313--2329, 1994. \href{https://doi.org/10.1016/0020-7683(94)90154-6}{doi:10.1016/0020-7683(94)90154-6}.
\bibitem{coulais2016} C. Coulais, E. Teomy, K. de Reus, Y. Shokef, and M. van Hecke, ``Combinatorial design of textured mechanical metamaterials,'' \emph{Nature}, vol. 535, pp. 529--532, 2016. \href{https://doi.org/10.1038/nature18960}{doi:10.1038/nature18960}.
\bibitem{buckmann2014} T. B\"uckmann, M. Thiel, M. Kadic, R. Schittny, and M. Wegener, ``An elasto-mechanical unfeelability cloak made of pentamode metamaterials,'' \emph{Nature Communications}, vol. 5, 4130, 2014. \href{https://doi.org/10.1038/ncomms5130}{doi:10.1038/ncomms5130}.
\bibitem{buckmann2015} T. B\"uckmann, M. Kadic, R. Schittny, and M. Wegener, ``Mechanical cloak design by direct lattice transformation,'' \emph{Proceedings of the National Academy of Sciences}, vol. 112, no. 16, pp. 4930--4934, 2015. \href{https://doi.org/10.1073/pnas.1501240112}{doi:10.1073/pnas.1501240112}.
\bibitem{bessa2019} M. A. Bessa, P. Glowacki, and M. Houlder, ``Bayesian machine learning in metamaterial design: Fragile becomes supercompressible,'' \emph{Advanced Materials}, vol. 31, no. 48, 1904845, 2019. \href{https://doi.org/10.1002/adma.201904845}{doi:10.1002/adma.201904845}.
\bibitem{wooldridge1995} M. Wooldridge and N. R. Jennings, ``Intelligent agents: Theory and practice,'' \emph{The Knowledge Engineering Review}, vol. 10, no. 2, pp. 115--152, 1995. \href{https://doi.org/10.1017/S0269888900008122}{doi:10.1017/S0269888900008122}.
\bibitem{gaia2000} M. Wooldridge, N. R. Jennings, and D. Kinny, ``The Gaia methodology for agent-oriented analysis and design,'' \emph{Autonomous Agents and Multi-Agent Systems}, vol. 3, pp. 285--312, 2000. \href{https://doi.org/10.1023/A:1010071910869}{doi:10.1023/A:1010071910869}.
\bibitem{corkill1991} D. D. Corkill, ``Blackboard systems,'' \emph{AI Expert}, vol. 6, no. 9, pp. 40--47, 1991. \href{https://dancorkill.com/pubs/ai-expert.pdf}{Author's unabridged article}.
\bibitem{whitley1999} D. Whitley, S. Rana, and R. B. Heckendorn, ``The island model genetic algorithm: On separability, population size and convergence,'' \emph{Journal of Computing and Information Technology}, vol. 7, no. 1, pp. 33--47, 1999. \url{https://hrcak.srce.hr/150198}.
\bibitem{krasnogor2005} N. Krasnogor and J. Smith, ``A tutorial for competent memetic algorithms: Model, taxonomy, and design issues,'' \emph{IEEE Transactions on Evolutionary Computation}, vol. 9, no. 5, pp. 474--488, 2005. \href{https://doi.org/10.1109/TEVC.2005.850260}{doi:10.1109/TEVC.2005.850260}.
\bibitem{burke2013} E. K. Burke, M. Gendreau, M. Hyde, G. Kendall, G. Ochoa, E. \"Ozcan, and R. Qu, ``Hyper-heuristics: A survey of the state of the art,'' \emph{Journal of the Operational Research Society}, vol. 64, no. 12, pp. 1695--1724, 2013. \href{https://doi.org/10.1057/jors.2013.71}{doi:10.1057/jors.2013.71}.
\bibitem{baluja1994} S. Baluja, ``Population-based incremental learning: A method for integrating genetic search based function optimization and competitive learning,'' Technical Report CMU-CS-94-163, Carnegie Mellon University, 1994. \href{https://publications.ri.cmu.edu/population-based-incremental-learning-a-method-for-integrating-genetic-search-based-function-optimization-and-competitive-learning}{Institutional publication record}.
\bibitem{mladenovic1997} N. Mladenovi\'c and P. Hansen, ``Variable neighborhood search,'' \emph{Computers \& Operations Research}, vol. 24, no. 11, pp. 1097--1100, 1997. \href{https://doi.org/10.1016/S0305-0548(97)00031-2}{doi:10.1016/S0305-0548(97)00031-2}.
\bibitem{kirkpatrick1983} S. Kirkpatrick, C. D. Gelatt, Jr., and M. P. Vecchi, ``Optimization by simulated annealing,'' \emph{Science}, vol. 220, no. 4598, pp. 671--680, 1983. \href{https://doi.org/10.1126/science.220.4598.671}{doi:10.1126/science.220.4598.671}.
\bibitem{virtanen2020} P. Virtanen et al., ``SciPy 1.0: fundamental algorithms for scientific computing in Python,'' \emph{Nature Methods}, vol. 17, pp. 261--272, 2020. \href{https://doi.org/10.1038/s41592-019-0686-2}{doi:10.1038/s41592-019-0686-2}.
\bibitem{bell2023} N. Bell, L. N. Olson, J. Schroder, and B. Southworth, ``PyAMG: Algebraic Multigrid Solvers in Python,'' \emph{Journal of Open Source Software}, vol. 8, no. 87, p. 5495, 2023. \href{https://doi.org/10.21105/joss.05495}{doi:10.21105/joss.05495}.
\end{thebibliography}
\end{document}